\documentclass[twocolumn]{aastex631}

\newcommand\calib{-4.362}
\newcommand\calibstat{0.033}
\newcommand\calibsys{0.045}

\usepackage{amsmath}

\begin{document}


\title{The TRGB$-$SBF Project. I. \\ A Tip of the Red Giant Branch Distance to the Fornax Cluster with JWST} 


\author[0000-0002-5259-2314]{Gagandeep S. Anand}
\affiliation{Space Telescope Science Institute, 3700 San Martin Drive, Baltimore, MD 21218, USA}

\author[0000-0002-9291-1981]{R. Brent Tully}
\affiliation{Institute for Astronomy, University of Hawaii, 2680 Woodlawn Drive, Honolulu, HI 96822, USA}

\author[0000-0001-5487-2494]{Yotam Cohen}
\affiliation{Space Telescope Science Institute, 3700 San Martin Drive, Baltimore, MD 21218, USA}

\author[0000-0001-9110-3221]{Dmitry I. Makarov}
\affiliation{Special Astrophysical Observatory of the Russian Academy of Sciences, Nizhnij Arkhyz, Karachay-Cherkessia 369167, Russia}

\author[0000-0003-0736-7609]{Lidia N. Makarova}
\affiliation{Special Astrophysical Observatory of the Russian Academy of Sciences, Nizhnij Arkhyz, Karachay-Cherkessia 369167, Russia}

\author[0000-0001-8762-8906]{Joseph B. Jensen}
\affiliation{Department of Physics, Utah Valley University, 800 W. University Parkway, Orem, UT 84058, USA}

\author[0000-0002-5213-3548]{John P. Blakeslee}
\affiliation{NSF's NOIRLab, 950 N Cherry Ave, Tucson, AZ 85719, USA}

\author[0000-0003-2072-384X]{Michele Cantiello}
\affiliation{INAF $–$ Astronomical Observatory of Abruzzo, Via Maggini, 64100, Teramo, Italy}

\author[0000-0002-5514-3354]{Ehsan Kourkchi}
\affiliation{Institute for Astronomy, University of Hawaii, 2680 Woodlawn Drive, Honolulu, HI 96822, USA}
\affiliation{Department of Physics, Utah Valley University, 800 W. University Parkway, Orem, UT 84058, USA}

\author[0000-0002-5577-7023]{Gabriella Raimondo}
\affiliation{INAF $–$ Astronomical Observatory of Abruzzo, Via Maggini, 64100, Teramo, Italy}


\begin{abstract}

Differences between the local value of the Hubble constant measured via the distance ladder versus the value inferred from the cosmic microwave background with the assumption of the standard $\Lambda$CDM model have reached over 5$\sigma$ significance. To determine if this discrepancy is due to new physics or more mundane systematic errors, it is essential to remove as many sources of systematic uncertainty as possible by developing high-precision distance ladders that are independent of the traditional Cepheid and Type Ia supernovae route. Here we present JWST observations of three early-type Fornax Cluster galaxies, the first of fourteen observations from a Cycle 2 JWST program. 
Our modest integration times allow us to measure highly precise tip of the red giant branch (TRGB) distances, and will also be used to perform measurements of Surface Brightness Fluctuations (SBF). From these three galaxies, we determine an average TRGB distance modulus to the Fornax Cluster of $\mu$ = 31.424 $\pm$ 0.077~mag, or D = 19.3 $\pm$ 0.7~Mpc. With eleven more scheduled observations in nearby elliptical galaxies, our program will allow us set the zero point of the SBF scale to better than 2$\%$ for more distant measurements, charting a path towards a high-precision measurement of $H_{0}$ that is independent of the traditional Cepheid-SN~Ia distance ladder.

\end{abstract}


\section{Introduction} \label{sec:intro}

For at least a half a century, the Virgo Cluster has played an important role in the establishment of the extragalactic distance scale~\citep{Sandage1976}. However, the depth of the virializing region of this cluster, about $\pm10\%$ of its distance, is an appreciated hindrance (increasing the scatter in the mean distance to the cluster), made more difficult by the ongoing accretion of new members~\citep{Tully1984,2024arXiv240316235C}. The Fornax Cluster has played a more limited role, but it is more compact with a depth about $\pm5\%$ of its distance and little evidence of recent accretion. Fornax is only slightly more distant than Virgo;  surface brightness fluctuation (SBF) measurements for 43 galaxies in Fornax and 91 galaxies in the Virgo vicinity using the Advanced Camera for Surveys (ACS) on Hubble Space Telescope (HST), imply a Fornax/Virgo distance ratio $d_F/d_V = 1.21\pm0.02$~\citep{2009ApJ...694..556B}.

In the most recent compilation of Cosmicflows-4 distances~\citep{Tully2023}, distances are given for 52 galaxies associated with the Fornax Cluster, including 4 hosts of Type~Ia supernovae (NGC\,1365, NGC\,1316, NGC\,1404, and NGC\,1380). The first three of these also have distance measurements via the tip of the red giant branch (TRGB) method. Two Fornax galaxies have a distance determination from Cepheid variable stars (NGC\,1326A and NGC\,1365), the latter of which is a SN Ia host galaxy. The TRGB and Cepheid distance moduli for NGC\,1365 agree within 0.02 mag, or 1\% in distance. The five TRGB and Cepheid distances to four Fornax galaxies on the Cosmicflows-4 zero point scale have an average distance of $18.0\pm1.3$~Mpc. The scatter of 7\% is comparable to the depth of the cluster.

TRGB and Cepheid observations are a challenge at the distance of the Fornax Cluster with HST, requiring many HST orbits~\citep{2018ApJ...866..145H,2021ApJ...915...34H, 2022ApJ...934L...7R}. However, as will be demonstrated here, TRGB measurements can be made with high accuracy with much shorter integrations with JWST at the distance of this cluster. In this paper we report on observations of three galaxies in the cluster: NGC\,1404 with a prior HST TRGB measurement and host to two Type Ia supernovae~\citep{2021ApJ...915...34H,2022ApJ...932...15A}, NGC\,1380 that hosted a single Type Ia supernova~\citep{1993ApJ...415..589K}, and NGC\,1399, the central dominant galaxy in the cluster.

\begin{figure}
\epsscale{1.1}
\plotone{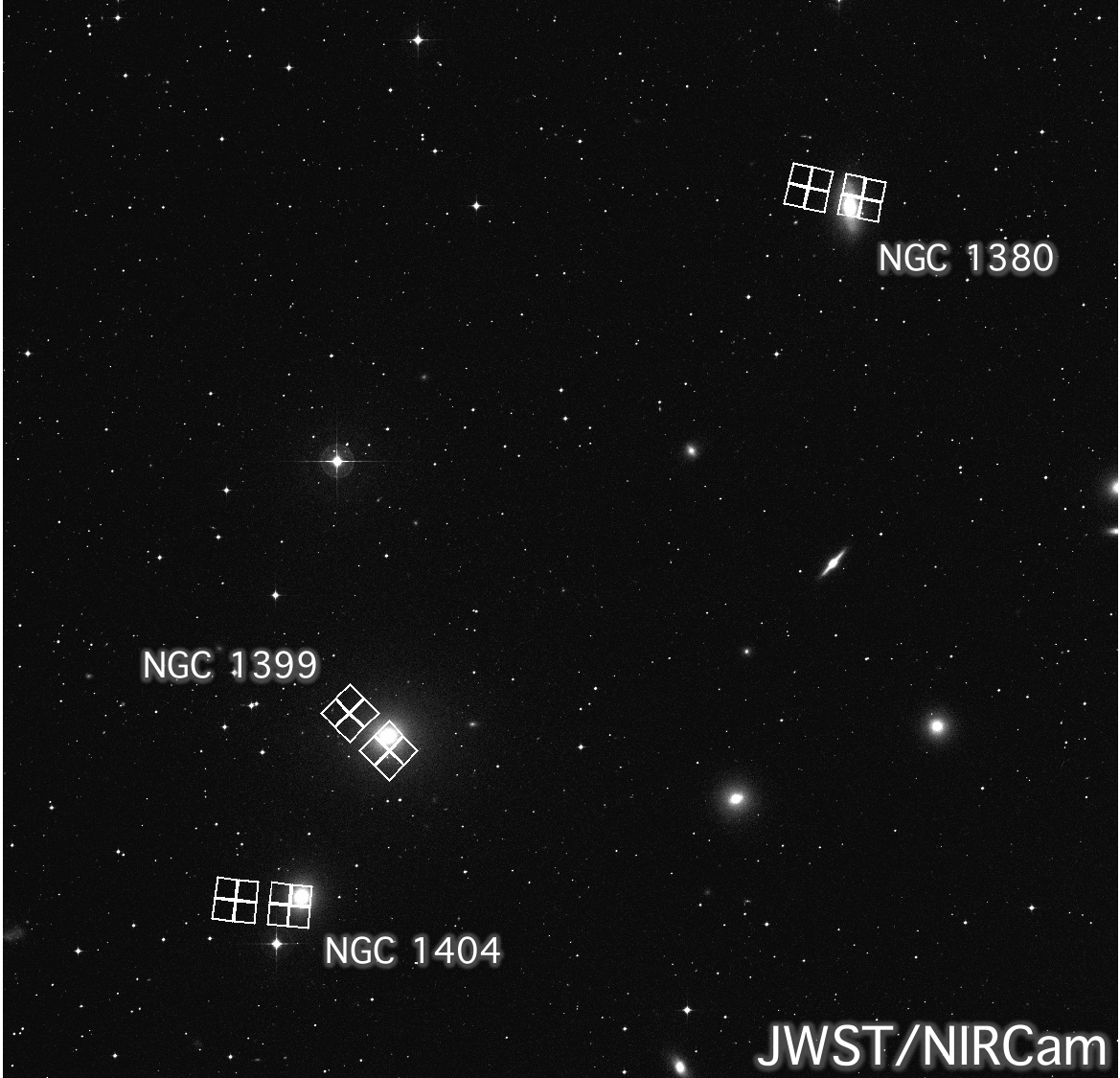}
\caption{Footprints of our JWST/NIRCam visits in the Fornax Cluster, overlaid on a one square degree image from the Digitized Sky Survey.}
\label{fig:footprints}
\end{figure}

Of equal interest to our program is the establishment of a robust zero-point calibration for the SBF methodology. The SBF technique with JWST holds great promise in measuring distances out into the Hubble flow~\citep{2021ApJ...911...65B, 2023arXiv230703116C}, allowing for an alternative to Type Ia supernovae as the final rung on the distance ladder~\citep{2022ApJ...934L...7R}, and a means to confirm or deny the $>$5$\sigma$ Hubble Tension between existing local distance ladder and cosmic microwave background derived measurements \citep{2021arXiv210301183D}. The old, gas-poor systems appropriate for SBF measurements do not contain significant numbers of Cepheids, requiring calibration, instead, with the TRGB. Aside from the peculiar galaxy Centaurus\,A and the obscured Maffei\,I, there are no giant elliptical galaxies within the 10~Mpc limit available for single orbit TRGB measurements with HST (a couple lie just beyond 10~Mpc in the Leo Group but already required multiple HST orbit observations for sufficient signal). As a consequence, TRGB distances have played only a minor role in establishing the SBF scale~\citep{Blakeslee2021}. With JWST, the very high point-source signal/noise (S/N) acquired with moderate exposures for galaxies within 20~Mpc provide an abundance of candidates to establish a TRGB-SBF linkage. Here we provide TRGB measurements to the three Fornax Cluster galaxies from our program and a first look at SBF measurements with JWST. 

\section{Data} \label{sec:data}

\begin{figure*}
\epsscale{1.15}
\plotone{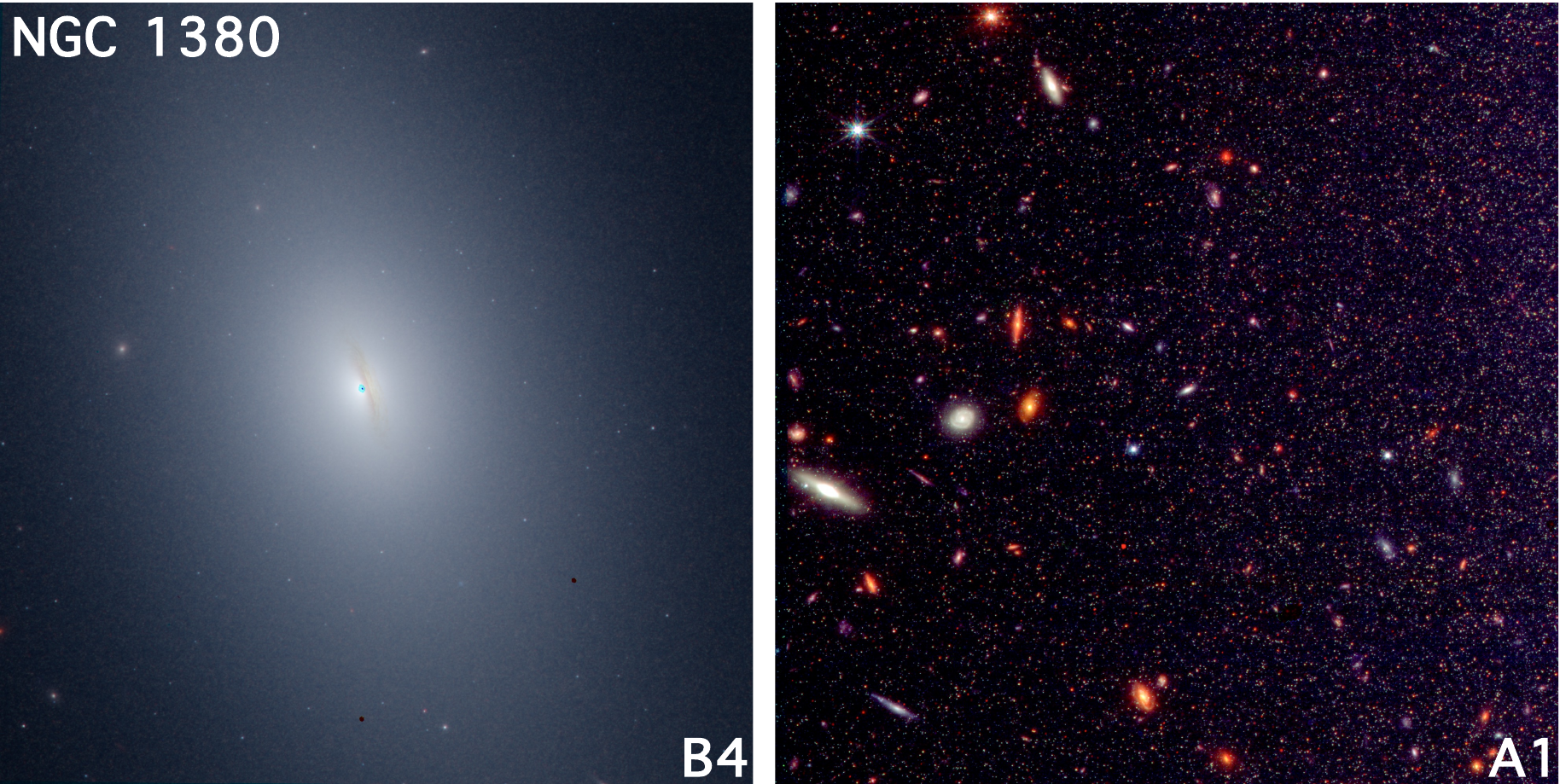}
\caption{Color images of the B4 and A1 chips from our NGC~1380 observations, generated from $F090W$/$F150W$/$F356W$ imaging. The elongated spatial coverage of NIRCam allows us to capture a large dynamic range of surface brightnesses within a single visit, facilitating simultaneous measurements of Surface Brightness Fluctuations and the Tip of the Red Giant Branch.}
\label{fig:n1380color}
\end{figure*}

The three galaxies presented in this work are the first of 14 targets of Cycle~2 JWST program GO-3055~\citep{2023jwst.prop.3055T} to link the TRGB and SBF scales through simultaneous observations tailored to each method. The split fields of the JWST NIRCam detector \citep{2023PASP..135f8001G, 2023PASP..135b8001R} are ideal for the experiment. One field covers the high surface brightness core of a galaxy for the SBF measurement, while the other field extends to the galactic halo where the TRGB measurement can be made with minimal confusion from crowding. SBF can also be measured in the peripheral fields, providing a direct overlap with TRGB. The primary observations are made in the $F090W$ and $F150W$ filters. $F090W$ is optimal for TRGB and SBF because of the minimal variance of RGB magnitudes at lower metallicities~\citep{2024ApJ...966...89A}. $F150W$ is another good choice for SBF because the fluctuation signal is very bright at this wavelength for the old, metal-rich populations of elliptical galaxies, making SBF measurable to distances beyond 100 Mpc \citep{2015ApJ...808...91J}. Simultaneous long-wavelength observations are made in both $F277W$ and $F356W$, although we do not make use of these filters in this work. The integration times are set by the requirements of the TRGB observations. The SBF signal is abundantly strong in the same exposure time. The deep multi-wavelength imaging of the selected galaxies is also useful for many other ancillary science cases.

\begin{deluxetable}{ccccc}[t]
\tabletypesize{\small}
\tablewidth{0pt}
\tablehead{
\colhead{Date} & \colhead{Target} & \colhead{Obs. ID} & \colhead{Filter} & \colhead{Time [s]}}
\startdata
2023-10-12 & NGC 1399 &  o002\_t002 &  F090W &   7730.474 \\
2023-10-12 & NGC 1399 &  o002\_t002 &  F150W &   1932.619 \\
2023-11-09 & NGC 1380 &  o001\_t001 &  F090W &   7730.474 \\
2023-11-09 & NGC 1380 &  o001\_t001 &  F150W &   1932.619 \\
2023-11-14 & NGC 1404 &  o003\_t003 &  F090W &   7730.474 \\
2023-11-14 & NGC 1404 &  o003\_t003 &  F150W &   1932.619 \\
\enddata
\caption{Observation log for the data used in this paper. All the images were processed with \texttt{jwst\_1140.pmap} or \texttt{jwst\_1147.pmap} context versions, which include the NIRCam flat-field and flux calibration updates introduced in \texttt{jwst\_1125.pmap} and \texttt{jwst\_1126.pmap}, respectively. The name of all observation IDs start with ``jw03055-" and end with ``\_nircam\_clear-f090w" or ``\_nircam\_clear-f150w". The times given are the actual measurement times (\texttt{TMEASURE}), and not the total exposure times (\texttt{EFFEXPTM}), which include some overhead. \label{tb:obs}}
\end{deluxetable}

Table~\ref{tb:obs} provides a summary of the data we use in this paper. All the data was reduced with context versions\footnote{See \url{https://jwst-crds.stsci.edu/display_all_contexts/} for more information on each version.} \texttt{1140.pmap} or \texttt{1147.pmap} (with no meaningful changes for NIRCam imaging between the two), and contain all reference file updates as of January 2024, including the latest updates to the photometric zeropoints, provided in \texttt{1126.pmap}. The observational setup with NIRCam was identical for each of the three galaxies presented in this paper. For each $F090W$ observation, we used a \texttt{MEDIUM8} readout pattern with 7 groups/integration, 3 integrations/exposure, and four subpixel dithers. For each $F150W$ observation, we used a \texttt{SHALLOW4} readout pattern, with 10 groups/integration, 1 integration/exposure, and four subpixel dithers.

Footprints of our NIRCam observations are shown in Fig.~\ref{fig:footprints}, overlaid on a one square degree image from the Digitized Sky Survey. From this figure, the extended spatial NIRCam coverage is clear. To further highlight the dynamic range probed by NIRCam in a single pointing, we show color ($F090W+F150W+F356W$) images of individual chips from the NGC~1380 visit in Fig.~\ref{fig:n1380color}. The core of this galaxy is centered on the B4 chip; a small dust lane in the center is readily apparent. While individual stars cannot be resolved in the inner regions due to crowding, a few globular clusters are visible scattered throughout as marginally resolved sources. Some background galaxies can also be seen. The two chips furthest from the core of the galaxy are A1 and A2, and we show a color image from the former. There are thousands of giant branch stars which populate this frame. Importantly, they can be easily separated from one another, due to the significantly lower surface brightness at this radius. A diverse population of background galaxies is also visible.

\section{Data Reduction} \label{sec:redux}

\begin{figure*}
\epsscale{1.15}
\plotone{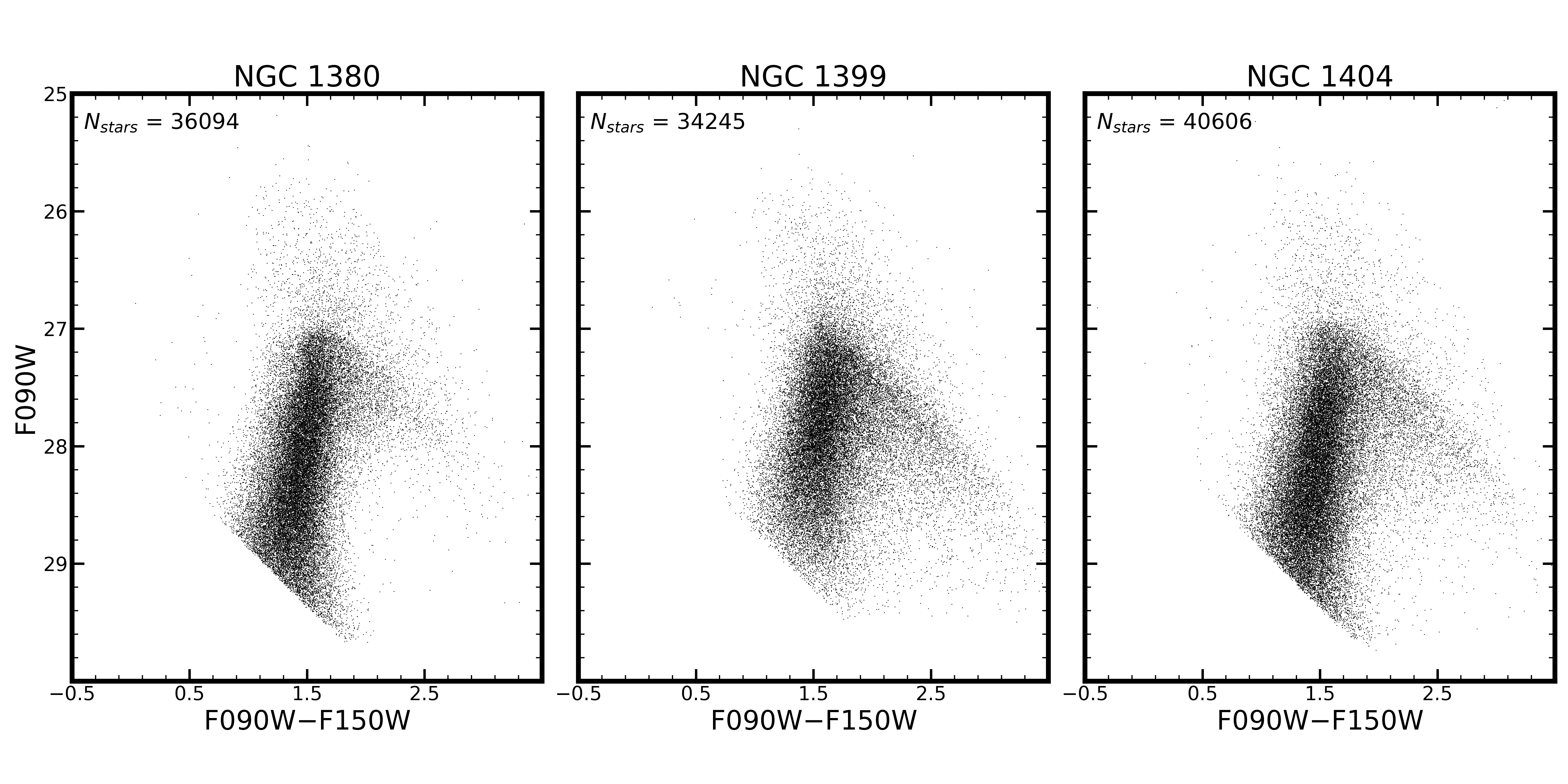}
\caption{Color-magnitude diagrams (uncorrected for foreground extinction) for three Fornax Cluster galaxies in our program. We only show stars from the two furthest of the eight SW chips (A1+A2). The TRGB feature is visible at approximately $m = 27$~mag, or $D = 19$~Mpc for each galaxy. For NGC\,1399, we only show the left-most quarter of each chip. The number of stars in each CMD is displayed in the upper-left for each target.}
\label{fig:triCMD}
\end{figure*}

Our JWST resolved star photometry procedure generally follows that as outlined in other recent works~\citep{2023ApJ...956L..18R, 2024ApJ...962L..17R, 2024ApJ...966...89A}. We utilize the DOLPHOT PSF photometry package~\citep{2000PASP..112.1383D, 2016ascl.soft08013D}– specifically, we use the NIRCam module, which has been developed as part of the JWST Resolved Stellar Populations Early Release Science Program~\citep{Weisz2023, 2024ApJS..271...47W}. We use the latest major version of the software (and associated PSFs) available at the time of writing this paper, which is the update provided on February 4th, 2024. 

For our photometry, we use the recommended parameters provided in the recent work of \cite{2024ApJS..271...47W}. We perform our PSF photometry directly on the Stage 2 \texttt{cal} images, while using a Stage 3 \texttt{i2d} image from the JWST pipeline as the ``reference frame'' for mutual image alignment. We choose $F090W$ for the reference frame, as it has the highest angular resolution, and our observations are also the deepest in that filter. We filter our final object list based on the quality criteria presented in \cite{2024ApJ...966...89A}, which are based on the recommendations provided by the ERS program team in \cite{Warfield2023}. We only include sources which meet the following DOLPHOT criteria in both filters in our final object list (except for type, which is not a filter specific output): (1) Crowding $<$ 0.5; (2) $\mathrm{(Sharpness)^{2}}$ $\le$ 0.01; (3) Object Type $\le$ 2; (4) S/N $\ge$ 5; (5) Error Flag $\le$ 2. All of our PSF photometry is transformed to the Vega system (and not the Sirius--Vega variant\footnote{By default, DOLPHOT uses the Sirius--Vega system. This can be switched to the Vega system by switching the \texttt{filters\_vega.dat} $\rightarrow$ \texttt{filters.dat} in DOLPHOT's NIRCam installation folder. The difference between the two photometric systems is $\sim$0.04 mag in $F090W$, and $\sim$0.02 mag in $F150W$.}).

We opt to utilize the ``--etctime'' option in the DOLPHOT/NIRCam module, which rectifies the effective exposure time provided in the JWST image headers (\texttt{EFFEXPTM}) to use the appropriate measurement time (\texttt{TMEASURE}) for derivation of the underlying S/N. The use of this option does not significantly affect the flux measurements themselves (due to the underlying image units of surface brightness); instead it aims to provide more realistic S/N estimates for each individual source~\citep{2024ApJS..271...47W}.

We also performed artificial star experiments with DOLPHOT. One at a time, 100,000 artificial stars (per chip) of known magnitude and position are inserted and recovered. The spatial distribution of these artificial stars was set to match the distribution of genuine stars found in the image. The same quality filtering is applied to the artificial stars, allowing us to quantify the effects of photometric bias and incompleteness that may be present in our data.

Color-magnitude diagrams (CMDs) for each target can be seen in Fig.~\ref{fig:triCMD}. These CMDs are limited to the two SW NIRCam chips that are furthest from the galaxy's core (A1+A2). For NGC\,1399, we further restrict the points to just those that are in the furthest quarter of each chip, for visual clarity. In each CMD, the vast majority of resolved stars are those on the red-giant branch (RGB), with the TRGB visible at $m_{F090W} \simeq 27$~mag ($D = 19$~Mpc) for each target. The magnitude limit of the CMD for NGC\,1399 is $\sim0.3$~mag brighter than for the other two targets, likely due to crowding effects at the faint end (NGC\,1399 is the cD galaxy in the Fornax Cluster, and thus more extended). It can also be seen that the RGB contains stars of a vast range of metallicities; for our TRGB work, we will only use the lower metallicity stars, which have little-to-no variation in their magnitude when observed in $F090W$ (see the discussion in \S2 of \citealt{2024ApJ...966...89A}). A further description of the underlying stellar populations can be found in \S4.6. 

On a broader note, the CMDs from our current program also highlight the great promise of performing TRGB measurements and general resolved stellar population science with JWST for galaxies well past 20~Mpc. It is widely accepted that 1 magnitude below the TRGB is needed for rigorous TRGB measurements~\citep{ 1995AJ....109.1645M, 2006AJ....132.2729M, 2009ApJ...690..389M, 2021AJ....162...80A}. This requirement of having the TRGB be one magnitude brighter than the S/N = 5 photometric limit means that our current program design would allow a secure measurement of the TRGB out to $m_{F090W} = 28$~mag, $\mu = 32.4$~mag, or $D = 30$~Mpc without any further increase in exposure time (total visit times of $\sim4.5$~hours). 
It can also be seen from Fig.~\ref{fig:triCMD} that the current incompleteness ``wedge'' is driven by the limited amount of exposure time in $F150W$, and not $F090W$ (see Table~\ref{tb:obs} for our exposure times). A modest amount of increased exposure time ($\sim$1--2~hours) devoted to additional $F150W$ observations would allow us to reach $m_\mathrm{TRGB} = 28.8$~mag, $\mu = 33.2$~mag, or $D = 44$~Mpc (with total visit times of $\sim$6~hours). We note that our measured S/N values are considerably better than expected, when compared to the JWST Exposure Time Calculator (v3.0). This likely has to do with the ETC's underlying assumption of aperture photometry; PSF photometry can often provide much higher S/N photometry than aperture photometry, especially in the faint regime. As of the writing of this paper, there is no option for the ETC to provide S/N estimates assuming PSF photometry, limiting its application in cases like ours.


\section{TRGB Measurements} \label{sec:trgb}

We now walk through the process of measuring TRGB distances to our three Fornax Cluster galaxies. We use two main techniques for measuring the TRGB; the first involves the use of an edge-detection algorithm~\citep{1993ApJ...417..553L, 2019ApJ...885..141B}, while the second method requires fitting a model luminosity function to the observed data~\citep{mendez2002}, and directly takes into account the results of artificial star experiments~\citep{2006AJ....132.2729M}. 

For each target, we perform our measurements on each of the two chips that are furthest from the galaxy's core. Due to the observational setup for our program, these are the A1 and A2 chips for all three targets. We analyze the data for both chips separately, reducing any concerns regarding the chip-to-chip zeropoint offsets. We note that for most standard stars, these offsets are at the level of 0.00--0.03~mag\footnote{See the details provided at \url{https://jwst-docs.stsci.edu/jwst-data-calibration-considerations/jwst-calibration-uncertainties} for more information regarding calibration uncertainties.}. For NGC\,1399, we also limit our analysis to the least crowded quarter of each of the two chips (via a simple spatial cut on the x-axis) in an attempt to mitigate the effects of the (modest) crowding present (which is more of an issue for the edge-detection routine, as it does not employ the use of artificial stars). 

With both of our chosen measurement techniques, we also restrict our analysis to the color range of $0.5 \le F090W-F150W \le 1.75$~mag. The red end of the color range is chosen to limit the analysis before the TRGB becomes fainter with metal-rich stars, as recommended by \cite{2024ApJ...966...89A}. The precise blue color is less important, as this end is typically chosen to eliminate young stars, such as main-sequence stars or yellow supergiants (or very low metallicity red giants, which we do not find in our observations). The lack of these stellar populations in our massive, early-type galaxies allows us to choose a relatively broad blue color, allowing for greater sampling of the RGB beneath the tip. 

\begin{deluxetable}{lccc}[t]
\tabletypesize{\small}
\tablewidth{0pt}
\tablehead{
\colhead{Field} & \colhead{Sobel TRGB} & \colhead{ML TRGB}}
\startdata
NGC 1380$-$A1 & 27.06 $\pm$ 0.01 & 27.061 $\pm$ 0.017  \\ 
NGC 1380$-$A2 & 27.08 $\pm$ 0.02 & 27.081 $\pm$ 0.018  \\ 
NGC 1399$-$A1 & 27.11 $\pm$ 0.02 & 27.169 $\pm$ 0.023  \\ 
NGC 1399$-$A2 & 27.12 $\pm$ 0.04 & 27.190 $\pm$ 0.023  \\ 
NGC 1404$-$A1 & 27.03 $\pm$ 0.03 & 27.035 $\pm$ 0.020  \\ 
NGC 1404$-$A2 & 27.00 $\pm$ 0.02 & 27.026 $\pm$ 0.019  \\ \hline
\enddata
\caption{A summary of our TRGB measurements from both the Sobel edge-detection method and the maximum-likelihood model fits. The edge-detection results are only given to two decimal points, as the underlying luminosity function is binned at the 0.01 mag level.}
\label{tab:trgb-measurements}
\end{deluxetable}

\begin{figure*}
\epsscale{1.15}
\plotone{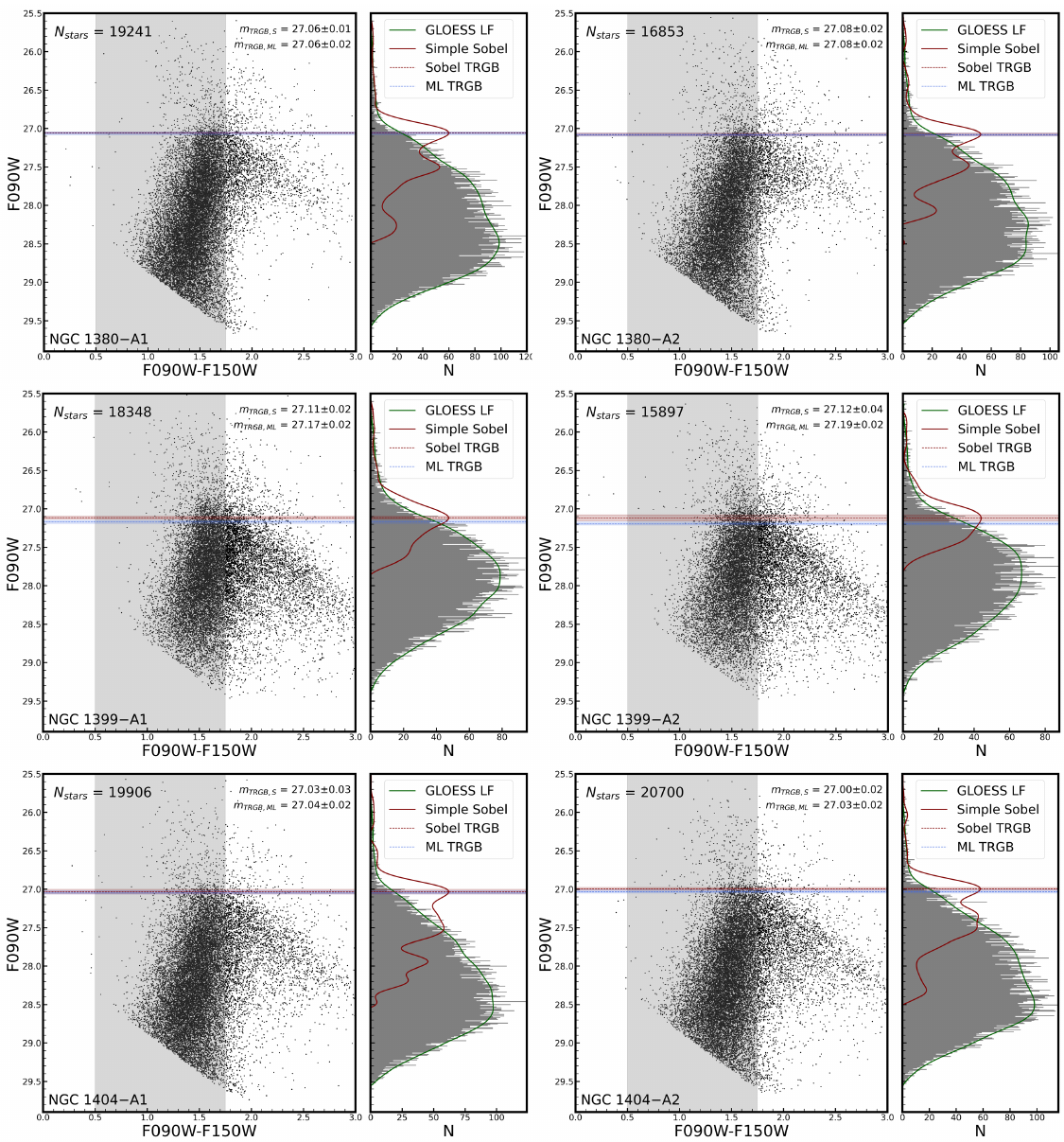}
\caption{Color-magnitude diagrams and edge-detection measurements for each of our six fields in three galaxies. The maximum-likelihood TRGB results are also plotted for comparison, but are nearly indistinguishable from the edge-detection results except for the NGC~1399 fields (which suffer from a modest amount of photometric bias). The grey-shaded region indicates the color range used for the TRGB measurements.}
\label{fig:mainTRGB}
\end{figure*}

\subsection{Edge-Detection}

To perform our baseline edge-detection measurements, we use the method outlined by \cite{2024ApJ...966...89A}, which is largely based on techniques first introduced by \cite{1993ApJ...417..553L, 1996ApJ...461..713S} and further developed by the Carnegie Chicago Hubble Program (CCHP; \citealt{2016ApJ...832..210B, 2019ApJ...882...34F}). To briefly summarize, we first restrict the analysis to the selected color range, and then marginalize over this color axis to generate a binned luminosity function with a bin-width of 0.01~mag. We then apply a GLOESS smoothing kernel with a smoothing scale of $\sigma_{s} = 0.10$~mag, similar to what is typically adopted in the literature\footnote{The effects of varying the smoothing kernel are typically at the level of $\sim$0-2$\%$ in distance, however we do not further explore these systematics in this work due to our eventual adoption of the maximum-likelihood derived measurements.}~\citep{2019ApJ...882...34F}. To detect the location of the TRGB, we employ a Sobel edge detection algorithm with a kernel of $[-1,0,1]$ (akin to a first derivative). We then assign the first prominent response as the brightness of the TRGB. 

The measurements are shown in Fig.~\ref{fig:mainTRGB}. The uncertainties are obtained via 1000 bootstrap resampling with replacement trials. We find good agreement between measurements from the two distinct chips for each galaxy, with the differences ranging from 0.01~mag (NGC\,1399) to 0.03~mag (NGC\,1404). These modest offsets are of the order of both (1) the potential current chip-to-chip offsets and (2) the uncertainties for each measurement, so it is difficult to ascertain their underlying nature at present. At this point, we do not combine photometry from the separate chips to avoid potentially skewing the results -- given that the brightest RGB stars will define the tip, outstanding zeropoint offsets would skew the measurements towards brighter values.

\begin{figure}
\epsscale{1.}
\plotone{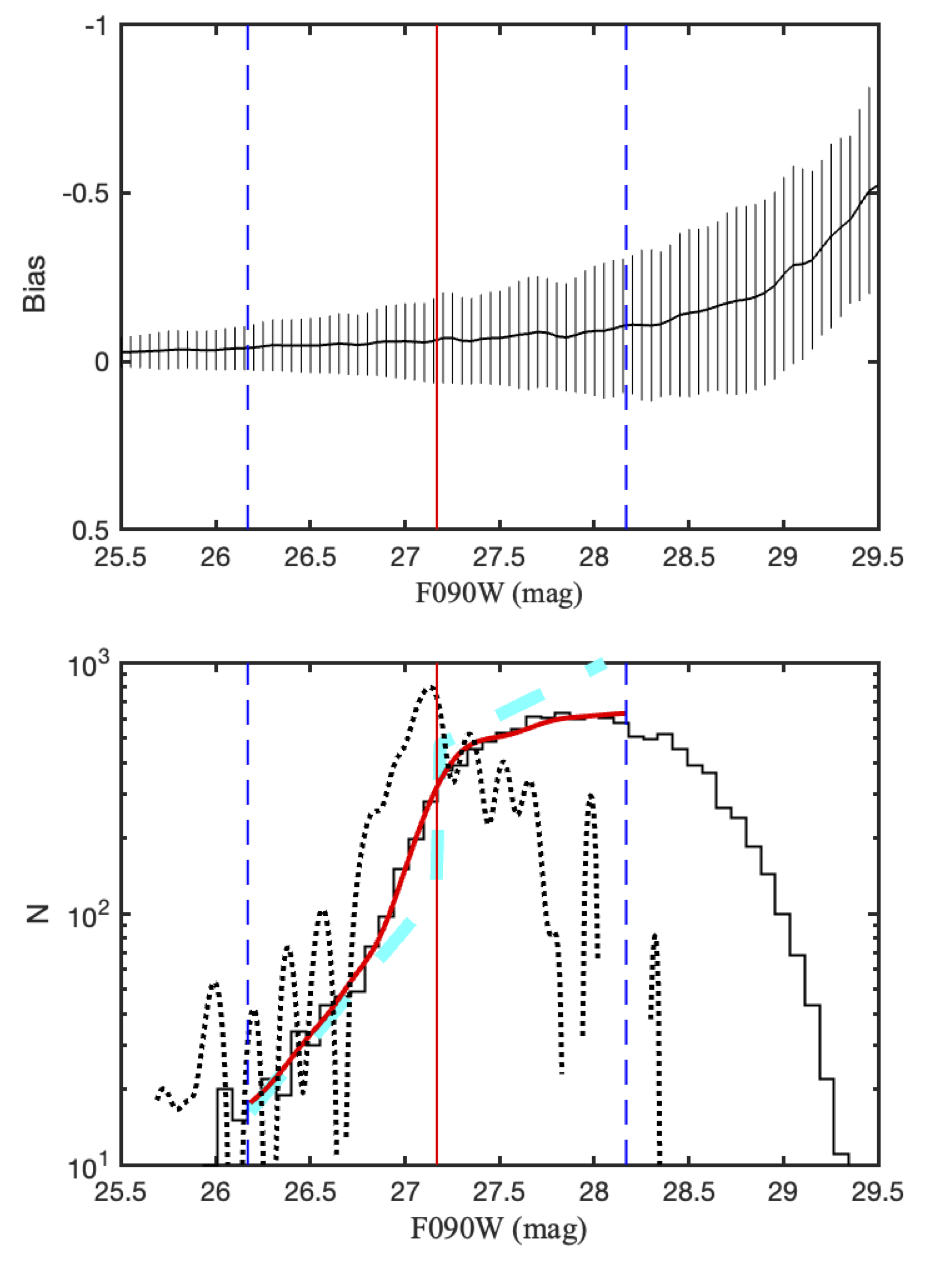}
\caption{\textbf{Top:} $F090W$ photometric bias curve for NGC~1399–A1, as a function of observed magnitude. The red vertical line denotes the TRGB, and the blue vertical lines indicate the magnitude range ($\pm$1 mag) used for the TRGB measurement procedure. The differences between the edge-detection and maximum-likelihood results for this galaxy seen in Fig.~\ref{fig:mainTRGB} are likely a result of the edge-detection method not accounting for photometric bias in this modestly crowded field. \textbf{Bottom:} Maximum-likelihood TRGB fit for NGC\,1399$-$A1. A binned version of the underlying luminosity function is shown as a histogram (the procedure itself does not require binning). The best-fit luminosity function is shown as the red line, after accounting for the measured photometric bias, completeness, and errors as determined from the artificial star experiments. The intrinsic underlying luminosity function with distinct RGB and AGB slopes is shown as the cyan line. For illustrative purposes, an edge-detection measurement is shown as the dotted line.}
\label{fig:ml-example}
\end{figure}

\begin{deluxetable}{lccc}[t]
\tabletypesize{\small}
\tablewidth{0pt}
\tablehead{
\colhead{Field} & \colhead{$m_{TRGB}$} & \colhead{$A_{F090W}$} & \colhead{$\mu$}}
\startdata
NGC 1380$-$A1 & 27.061 $\pm$ 0.017 & 0.021 & 31.387 $\pm$ 0.066 \\ 
NGC 1380$-$A2 & 27.081 $\pm$ 0.018 & 0.021 & 31.407 $\pm$ 0.066 \\ 
NGC 1399$-$A1 & 27.169 $\pm$ 0.023 & 0.016 & 31.500 $\pm$ 0.068 \\ 
NGC 1399$-$A2 & 27.190 $\pm$ 0.023 & 0.016 & 31.521 $\pm$ 0.068 \\ 
NGC 1404$-$A1 & 27.035 $\pm$ 0.020 & 0.014 & 31.368 $\pm$ 0.067 \\ 
NGC 1404$-$A2 & 27.026 $\pm$ 0.019 & 0.014 & 31.359 $\pm$ 0.067 \\ \hline
\enddata 
\caption{TRGB distances derived to each field, as determined from the maximum likelihood results.}
\label{tab:trgb-distances}
\end{deluxetable}

\subsection{Model LF Fitting}

For our model luminosity function TRGB measurements, we use the maximum likelihood \texttt{TRGBTOOL} described in \cite{2006AJ....132.2729M}. To briefly summarize, we adopt a theoretical luminosity function (LF) of the form:
\begin{equation}
\psi = \begin{cases}
  10^{\,a(m-m_\mathrm{TRGB})+b}, & \mbox{if } m \geq m_\mathrm{TRGB} \\ 
  10^{\,c(m-m_\mathrm{TRGB})},   & \mbox{if } m < m_\mathrm{TRGB},
\end{cases}
\end{equation}
where $m_\mathrm{TRGB}$ is the magnitude of the TRGB, $b$ denotes the strength of the RGB jump, and $a$ and $c$ are the slopes of the luminosity function on, respectively, the RGB and AGB sides of the jump. This theoretical LF is used to fit the observed data, while taking into account the photometric bias, completeness, and errors derived from our artificial star experiments. An initial estimate of the TRGB is provided via the results of the edge-detection analysis, and the fitting procedure is repeated until convergence. 

An example maximum-likelihood measurement is shown in Fig.~\ref{fig:ml-example} for NGC\,1399$-$A1. The binned histogram corresponds to the underlying luminosity function, although the procedure itself does not involve binning. The red line shows the best-fit model luminosity function to the observed data, including photometric effects derived via our artificial star experiments. The intrinsic underlying luminosity function (i.e.{} if the data were free from bias, errors, and incompleteness) is shown as the dashed cyan line. It can be seen from the mismatch between the cyan and red lines that the photometry for this target suffers from a modest amount of incompleteness at fainter magnitudes, likely due to the fact that this is the most crowded of our three galaxies. An edge-detection result is shown in the black dotted line, for comparison only. For all three of our galaxies, the maximum-likelihood results agree to within 0.02~mag for the two distinct chips. The results from the maximum-likelihood routine are given in Table~\ref{tab:trgb-measurements} and Fig.~\ref{fig:mainTRGB}.

\subsection{Methodological Comparisons}

There is very good agreement for the TRGB measurements of NGC\,1380 ($\Delta = 0.00$~mag) and NGC\,1404 ($\Delta = 0.02$~mag) with the Sobel and maximum-likelihood techniques. However, there is a larger difference for NGC\,1399 ($\Delta$ = 0.07 mag), which suffers from a moderate amount of crowding and hence photometric bias. In this case, it is likely that the edge-detection results are being biased brightwards by a selection of blended RGB stars, whereas the maximum-likelihood measurements take the observed bias into account. Indeed, Fig.~\ref{fig:ml-example} shows that differences between input and output magnitudes for artificial stars at the maximum-likelihood derived magnitude of the TRGB is $\sim$0.05~mag, similar to the level of offset between the two methods for this case. Given that some of our data suffers from a moderate amount of photometric bias, we proceed with the maximum-likelihood results (which account for these effects) for our final reported distances. We recommend that others who wish to perform accurate TRGB distance measurements go through similar checks of photometric bias if they wish to proceed with measurements from routines which do not take into account artificial star results.

\subsection{Distances}

To provide an absolute scaling for our TRGB measurements, we utilize the JWST/NIRCam TRGB calibration provided by~\cite{2024ApJ...966...89A}. They use data taken in the outer regions of the megamaser host NGC\,4258 to provide a direct calibration in our filter of interest ($F090W$). The baseline absolute calibration they provide is $M_\mathrm{TRGB}^{F090W} = \calib \pm \calibstat$~(stat) $\pm \calibsys$~(sys)~mag. They also note that this baseline calibration may be further tailored to individual measurement techniques: a value of $M_\mathrm{TRGB}^{F090W} = -4.347$~mag is more appropriate for techniques which employ the use of LF fitting with artificial stars (e.g.{} \citealt{2006AJ....132.2729M}) and take into account photometric bias, errors, and incompleteness. It has been shown that the model-LF fitting techniques can provide measurements that are up to $\sim$0.05 magnitudes fainter than their edge-detection counterparts \citep{2022ApJ...932...15A}, consistent with the 0.03 mag offset in their absolute JWST calibration \citep{2024ApJ...966...89A}.  We also note that several modest changes in the recommended DOLPHOT parameters may result in minor offsets in the final photometry, though \cite{2024ApJS..271...47W} note that the performance is comparable.\footnote{An updated NIRCam TRGB calibration with the new DOLPHOT parameters and PSF models, along with a NIRISS TRGB calibration will be provided in a future, separate work (Anand et al. in prep). Preliminary results from that work show that the net effect is insignificant ($<$0.01~mag)}. To account for any potential mismatches, we choose to include an additional 0.03~mag contribution to the overall systematic error term. The magnitude of this term is also provided by \cite{2024ApJ...966...89A}, who obtain this value from several other works in the literature which examine the systematic effects of changing DOLPHOT parameters on the output photometry \citep{2021ApJ...906..125J, 2023MNRAS.521.1532J}. Our adopted absolute calibration now becomes $M_\mathrm{TRGB}^{F090W} = -4.347 \pm 0.033$~(stat) $\pm 0.054$~(sys)~mag (see Table~3 in \citealt{2024ApJ...966...89A} for all of the included error terms). 

\begin{deluxetable}{cccc}[t]
\tabletypesize{\small}
\tablewidth{0pt}
\tablehead{
\colhead{Galaxy} & \colhead{JWST TRGB} & \colhead{HST TRGB} & \colhead{HST Cepheid}}
\startdata
NGC 1316  &                   & $31.35 \pm .10$ &                   \\
NGC 1326A &                   &                 & $31.009 \pm .100$ \\
NGC 1365  &                   & $31.40 \pm .09$ & $31.378 \pm .056$ \\
NGC 1380  & $31.397 \pm .066$ &                 &                   \\
NGC 1399  & $31.511 \pm .068$ &                 &                   \\
NGC 1404  & $31.364 \pm .067$ & $31.27 \pm .10$ &                   \\
\hline
Average  & $31.424 \pm .077$ & $31.34 \pm .07$ & $31.194 \pm .260$ \\
\enddata 
\caption{A comparison of distances to Fornax Cluster galaxies, from TRGB and Cepheid measurements.}
\label{tab:qual-distances}
\end{deluxetable}

Until now, we have not included the minor contribution from foreground extinction in our analysis. To tabulate $A_{F090W}$ for each target, we take $E(B-V)$ from \cite{2011ApJ...737..103S} and adopt $A_{F090W}/E(B-V) = 1.4156$~\citep{2023ApJ...956L..18R}. The corresponding extinction values are $A_{F090W} = 0.021$~mag for NGC\,1380, $A_{F090W} = 0.016$~mag for NGC\,1399, and $A_{F090W} = 0.014$~mag for NGC\,1404. After correcting the observed magnitudes for foreground extinction, we tabulate $\mu = m-M$ for each field separately, and note the results in Table~\ref{tab:trgb-distances}. The uncertainties on the individual distance moduli are tabulated from the quadrature sum of the uncertainty in the observed magnitude of the tip and the uncertainty in the absolute magnitude calibration of $\pm0.063$~mag. The former term is itself the quadrature sum of the uncertainty in the underlying TRGB measurement, and a conservative one-half the value of the foreground extinction. For our final distance moduli, we take the average of each of the two individual fields (without further averaging down of the errors), to find $\mu = 31.397 \pm 0.066$~mag for NGC\,1380, $\mu = 31.511 \pm 0.068$~mag for NGC\,1399, and $\mu = 31.364 \pm 0.067$~mag for NGC\,1404 (Table \ref{tab:qual-distances}).

\begin{figure}
\centering
\includegraphics[width=\columnwidth]{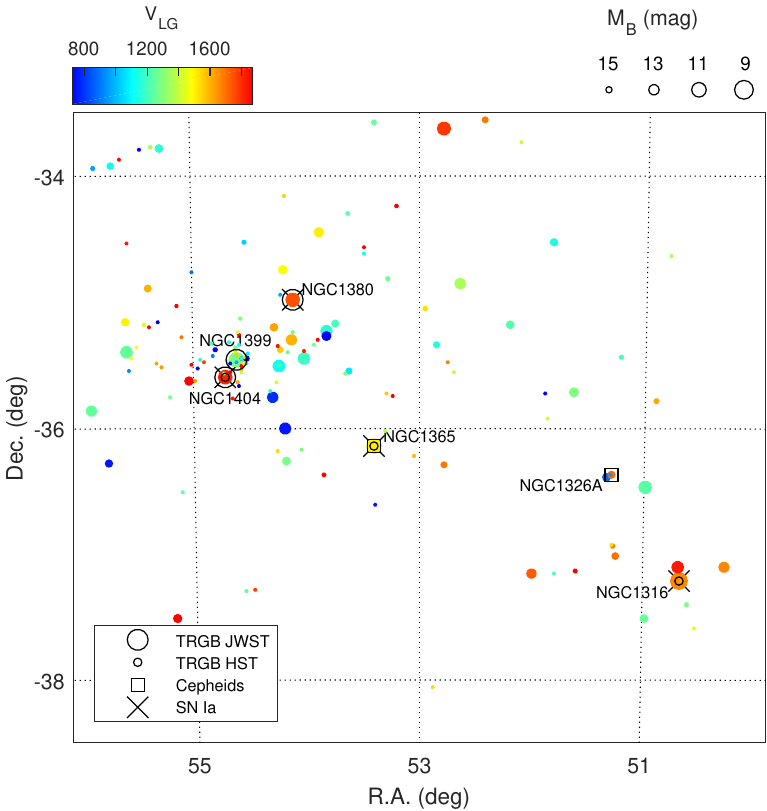}
\caption{Distribution of galaxies in the vicinity of the Fornax Cluster. Galaxies are colored according to their redshift in the Local Group reference frame. Big and small circles indicate objects with JWST and HST TRGB distances, respectively; squares mark two galaxies with Cepheid distances; and x-crosses stand for the four galaxies that hosted SN\,Ia.}
\label{fig:map}
\end{figure}

\begin{figure*}
\epsscale{1.1}
\plotone{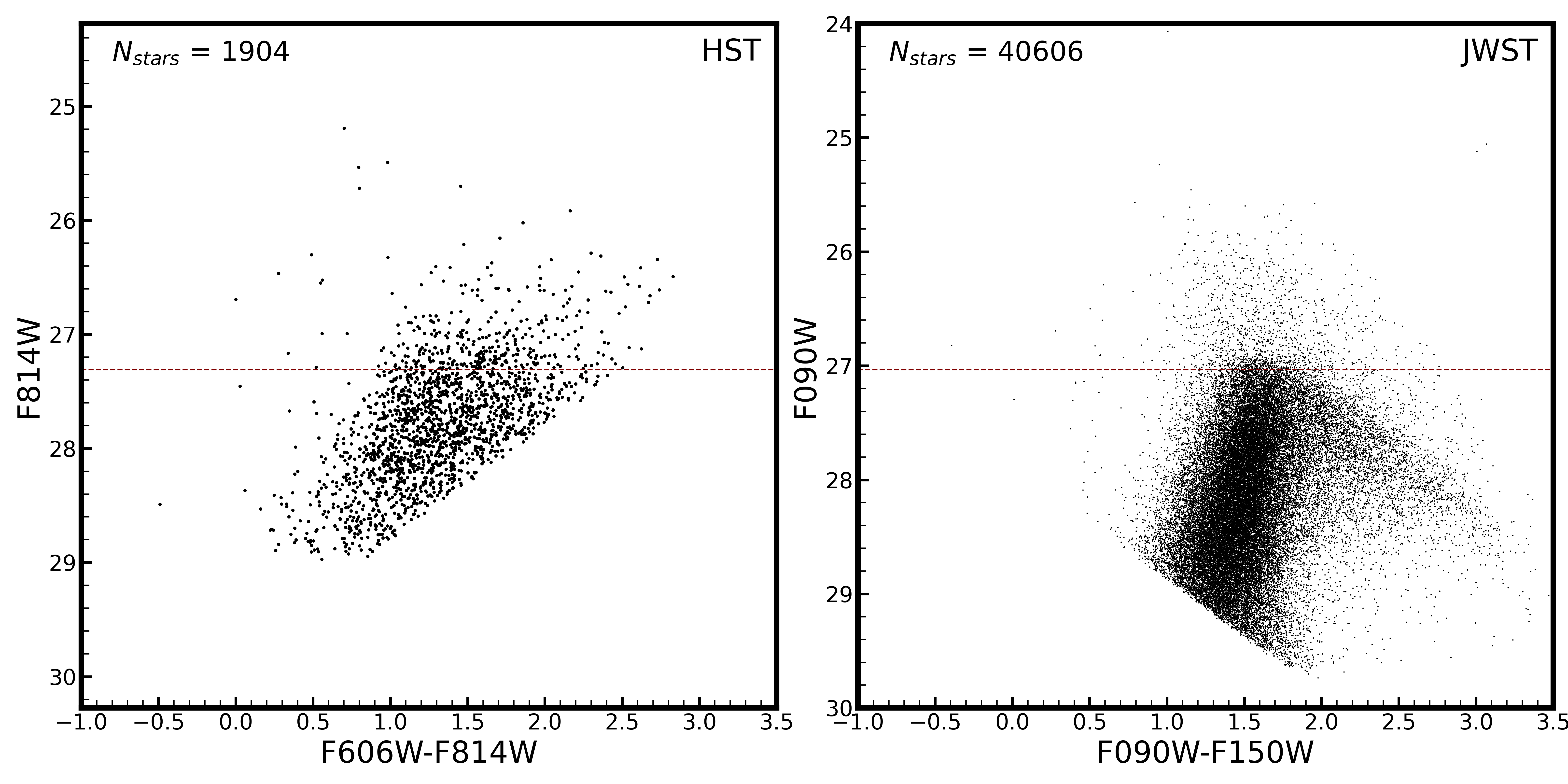}
\caption{Comparison of two CMDs (down to S/N = 5) for NGC\,1404 used to measure the TRGB distance to this galaxy. The left panel shows the reduction of the 34 orbit (crowding-limited) HST dataset, whereas the right panel shows the reduction of the 4.74 hour JWST dataset. The y-axes of the two panels are set to match at the measured levels of the TRGB (denoted with dashed red lines). Insets show the respective observing footprints overlaid on 12$'' \times$12$''$ DSS imaging, with the colored regions denoting the regions shown in the CMDs and used for the TRGB measurements.}
\label{fig:hst-jwst}
\end{figure*}

\textbf{}
\subsection{Comparison with Previous Distances}

There are distances for 52 galaxies in the Fornax Cluster in the Cosmicflows-4 compilation of distances~\citep{Tully2023}. The largest contribution (43 SBF distances) originate from the HST ACS Fornax Cluster Survey \citep{2009ApJ...694..556B}.  There are 18 (overlapping) Fundamental Plane entries, 8 Tully-Fisher distances, 4 Type Ia SNe, 3 TRGB measurements (from HST observations), and 2 Cepheid contributions. The Cosmicflows-4 ensemble averaged distance to the Fornax Cluster is $19.62 \pm 0.12$~Mpc with most of the weight of the result coming from the large SBF sample with low internal uncertainties (although \citealt{2009ApJ...694..556B} state a systematic uncertainty of 1.4~Mpc from their HST/ACS SBF data).

Given the significance of the statistics of the different distance measurement techniques, differing results depend sensitively on systematic errors arising from their absolute calibrations.  Table~\ref{tab:qual-distances} reports distance modulus measurements from TRGB and Cepheid Period-Luminosity studies. The averaged distances for the three columns are $19.3\pm0.7$~Mpc (3~JWST TRGB), $18.5\pm0.6$~Mpc (3~HST TRGB), and $17.3\pm2.2$~Mpc (2~HST Cepheid). The HST TRGB and Cepheid results in Table~\ref{tab:qual-distances} are on the absolute scale established in the Cosmicflows-4 study. The JWST TRGB results presented here are those based on the scale set by the distance to the megamaser host galaxy NGC\,4258, for which a geometrical distance good to 1.5\% has been measured \citep{2019ApJ...886L..27R}. The positions of galaxies in and around the Fornax Cluster are plotted in Fig.~\ref{fig:map}. Those with TRGB and Cepheid distances, and those that have hosted SN\,Ia, are highlighted. 

We note that the Cepheid distances to NGC~1326A \citep{2001ApJ...553...47F} and NGC~1365 \citep{2022ApJ...934L...7R} differ by nearly 20$\%$, significantly greater than the depth of the Fornax cluster. An initial HST Key project value of $\mu = 31.36 \pm 0.21$ mag for NGC~1326A was provided by \cite{1999ApJ...525...80P}, which greatly differs from the final measurements provided by the Key Project \citep{2001ApJ...553...47F} of $\mu = 31.01 \pm 0.10$ mag. A modern reduction of the underlying data might provide further insight into the apparent discrepancies.

As a particularly relevant example, we compare the color-magnitude diagrams down to S/N = 5 from HST and JWST observations in NGC\,1404 in Fig.~\ref{fig:hst-jwst}. The (crowding-limited) HST data was taken over 34 orbits (51 hours) as part of the Carnegie-Chicago Hubble Program~\citep{2018hst..prop15642F, 2021ApJ...915...34H}, and the photometry is taken from the Extragalactic Distance Database's CMDs/TRGB catalog~\citep{2021AJ....162...80A}. The JWST data is from the A1 and A2 chips as earlier presented in this work, and taken with 4.74 hours of total charged time. The incredible efficiency of our JWST observations is clear.

\subsection{Stellar Populations}

\begin{figure*}
    \centering
    \includegraphics[width=5.8cm]{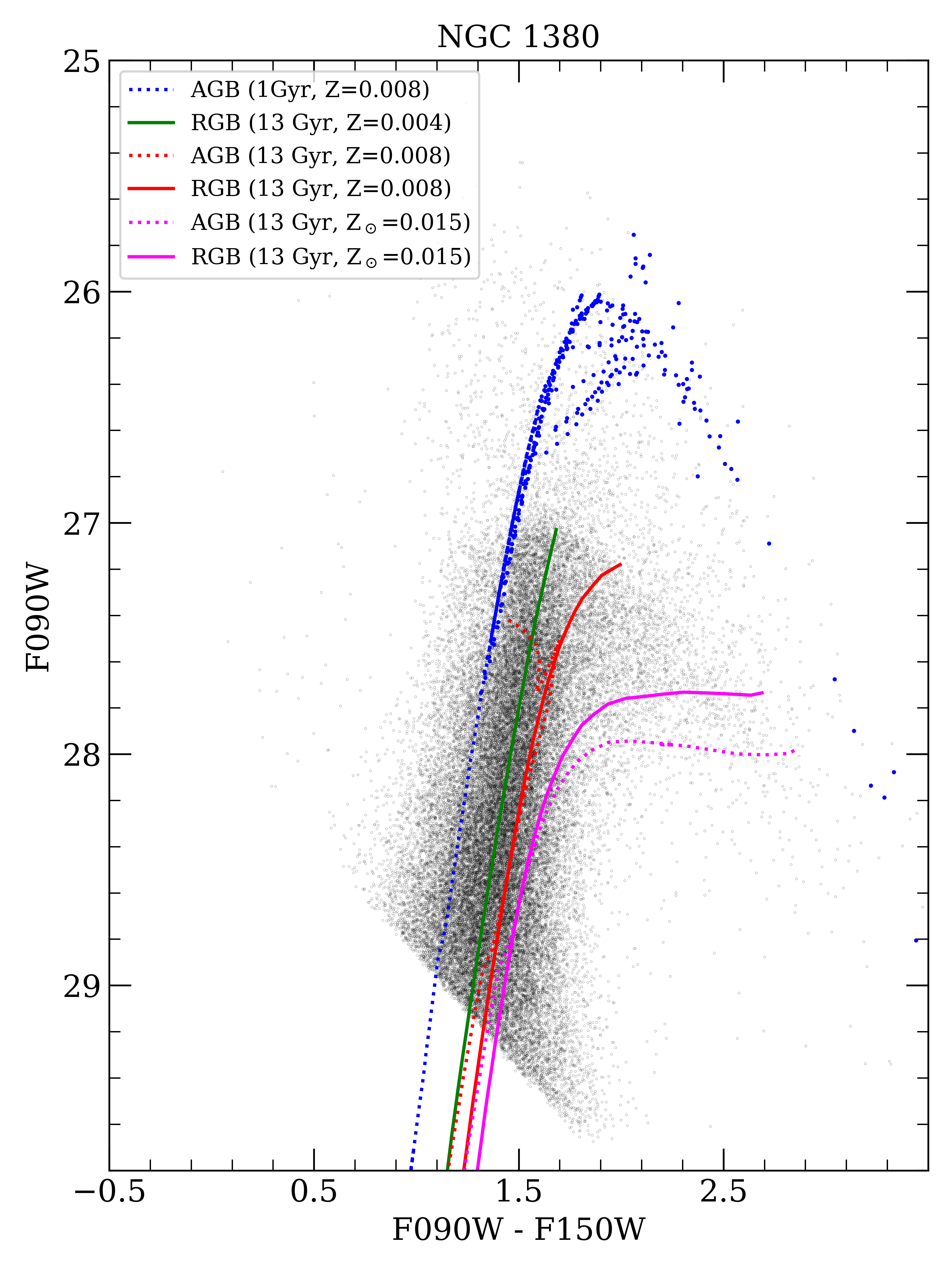}
    \includegraphics[width=5.8cm]{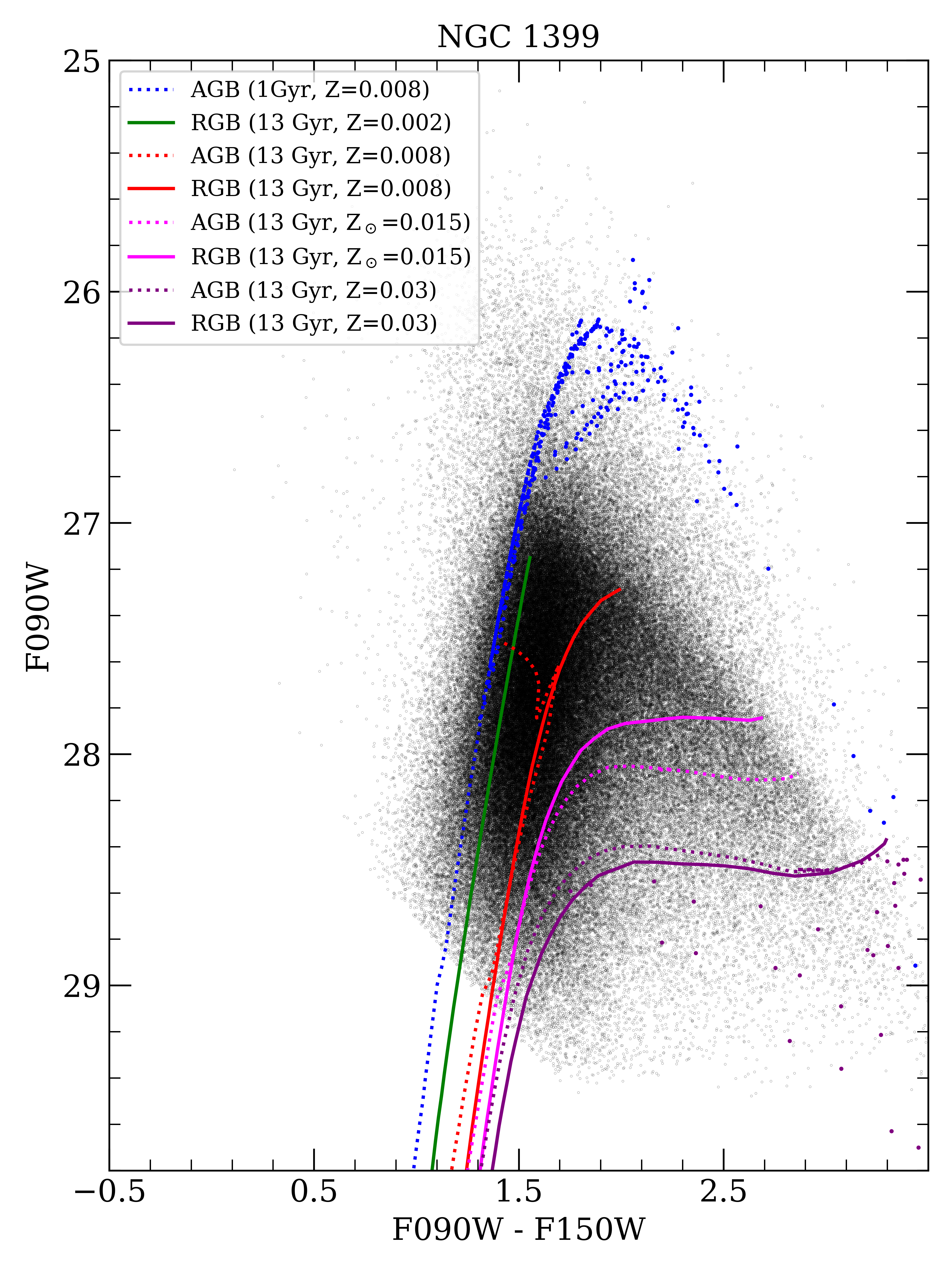}
    \includegraphics[width=5.8cm]{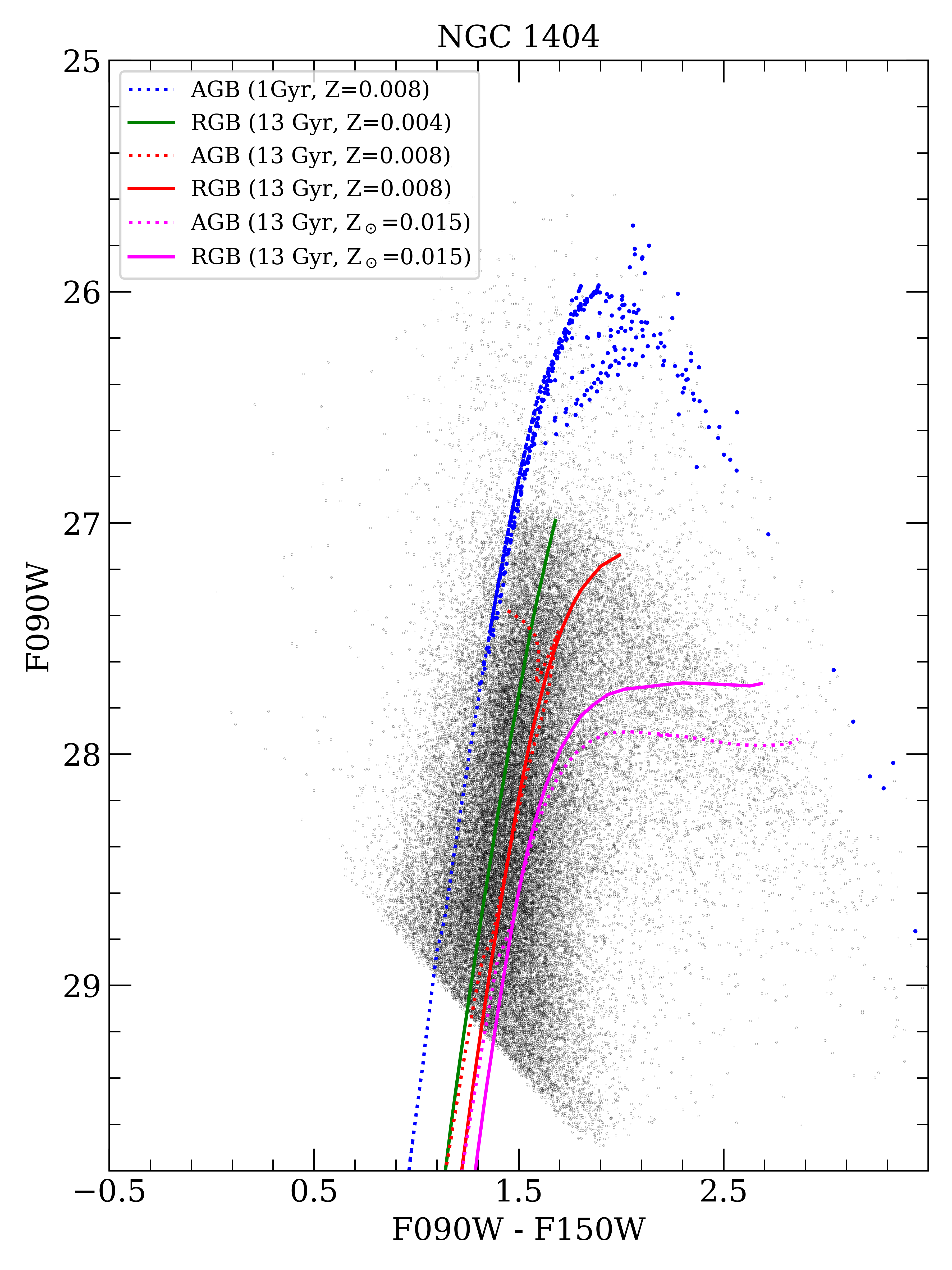}
    \caption{Color-magnitude diagrams of our selected NGC~1380, NGC~1399 and NGC~1404 fields, with the PARSEC stellar library isochrones overlaid. The isochrones were shifted according to the distances measured in this paper and Galactic extinction from \cite{2011ApJ...737..103S}.}
    \label{fig:isochrones}
\end{figure*}

In this section we briefly consider the resolved stellar populations of the fields under study. Figure~\ref{fig:isochrones} shows color-magnitude diagrams of these galaxies with PARSEC isochrones \citep{2012MNRAS.427..127B, 2013MNRAS.434..488M} of various ages and metallicities overplotted. The CMDs demonstrate the full A1+A2 fields, corresponding to outer regions of each galaxy $\sim$20~kpc from their centers. The central galaxy of the Fornax cluster is the giant elliptical NGC 1399. Spectral observations of NGC~1399 made with MUSE \citep{2018MNRAS.479.2443V} revealed a smooth decrease of the metallicity [Fe/H] from the center to the periphery, reaching a value slightly less than solar at the effective radius ($r_e$). To better understand the ages and metallicities of the resolved stars of NGC~1399, we take the PARSEC isochrones in the JWST/NIRCam system, using the distances derived in this work (Table~\ref{tab:qual-distances}). In general, the CMDs agree best with stellar population models with metallicity $Z$=0.008, somewhat less than solar metallicity ($Z_{\odot}$=0.015). Blue isochrones correspond to the E-AGB and TP-AGB evolutionary stages of 1 Gyr old stars. Thus, even in the outer regions of this galaxy, we can expect a number of stars with relatively young ages (about 1--2 Gyrs). Older RGB and AGB stars, up to the oldest at 13.7 Gyr, appear to form the bulk of the population. A fairly well populated ``tail'' stretching to the right and downwards is clearly visible, showing the presence of red giants and AGB stars of solar metallicity and higher.

Although the CMDs of all three galaxies are similar, the lenticular galaxy NGC~1380 shows fewer stars in its outer regions. While features similar to NGC~1399 can be seen, the range of colors of the resolved stellar populations and the size of the high metallicity ``tail'' appear smaller. We see that the metallicity of the bulk of the resolved stars is somewhat less than solar, while there are a smaller number of older and more metal rich stars (up to around solar metallicity) than in the larger elliptical galaxies. The CMD of the elliptical galaxy NGC~1404 is intermediate between NGC~1380 and NGC~1399 and exhibits essentially the same trends as the other two galaxies.


\section{The Next Step: Calibrating the SBF Distance Scale} \label{sec:sbf}

The simultaneous observation of both high and low surface brightness regions of the Fornax giant elliptical galaxies provides the data required to calibrate the SBF distance scale in multiple filters independent of any reliance on Cepheid variable stars or associations with spiral galaxies and young stellar populations, as has commonly been done for past SBF calibrations (e.g.{} \citealt{2009ApJ...694..556B}, \citealt{2015ApJ...808...91J}). The TRGB distances presented here, based on a geometric calibration of NGC~4258, provide the basis for a new SBF zero point calibration using old, Population II stars.

The SBF technique~\citep{1988AJ.....96..807T} is a statistical measurement of the spatial fluctuations in surface brightness of a population of stars, which is dominated by the luminous RGB stars in an old smooth elliptical galaxy. Nearer galaxies show a mottled appearance that fades into smoothness for more distant galaxies. Unlike the TRGB, SBF measurements \emph{do not require resolution of individual RGB stars,} and can be measured much further away, to distances beyond 100~Mpc with HST~\citep{2021ApJ...911...65B}, and potentially much further with JWST~\citep{2023arXiv230703116C}. The typical internal SBF distance errors (not including systematic uncertainty) with HST are less than 5\% per galaxy, putting SBF on the same statistical level as type Ia supernovae, and capable of reaching well into the Hubble flow. It is therefore advantageous to have a dual measurement of SBF and TRGB for the same galaxies using the same filters and cameras to calibrate SBF, eliminating as many sources of systematic uncertainty as possible. This will allow us to directly address the persistent conflict between local distance ladder measurements of the Hubble constant $H_0$ and the value inferred from $\Lambda$CDM cosmological models \citep{2020A&A...641A...6P}, sometimes referred to as the Hubble tension~\citep{2021arXiv210301183D}. This paper outlines the first steps towards our goal of establishing a distance ladder with no reliance on Cepheids or supernovae, with the ultimate goal of reaching a distance precision approaching 1\%.

The spatial variations in surface brightness in a galaxy arise from the Poisson statistics of the discrete stars in the stellar population. The pixel-to-pixel variation in the number of stars provides the primary SBF signal, which is convolved with the point-spread function of the observatory and normalized by the mean galaxy brightness. Additional contributions to the SBF signal are the white-noise floor from the detector readout and additional bumpiness from undetected globular clusters and background galaxies. In distant galaxies, the latter contributions are carefully measured and subtracted; in these relatively nearby JWST observations of Fornax galaxies we can easily identify and remove resolved clusters and galaxies. 

The SBF calibration also depends on properties of the stellar population \citep{1997ApJ...475..399T, 2009ApJ...694..556B}. Optical colors like $g{-}z$ are used to calibrate SBF absolute magnitudes. With these NIRCam observations, we can also use the SBF signal in multiple filters, particularly those at longer wavelengths, to better understand the homogeneity of the population and the contribution from AGB stars. A full understanding of the dependence of the JWST SBF on stellar population will require a larger range in galaxy properties than we have in these first three galaxies, so we do not present a calibration of SBF at the present time. We can, however, demonstrate the effectiveness of our SBF procedures using these images in multiple fields and filters.

\begin{figure}
\epsscale{1.0}
\plotone{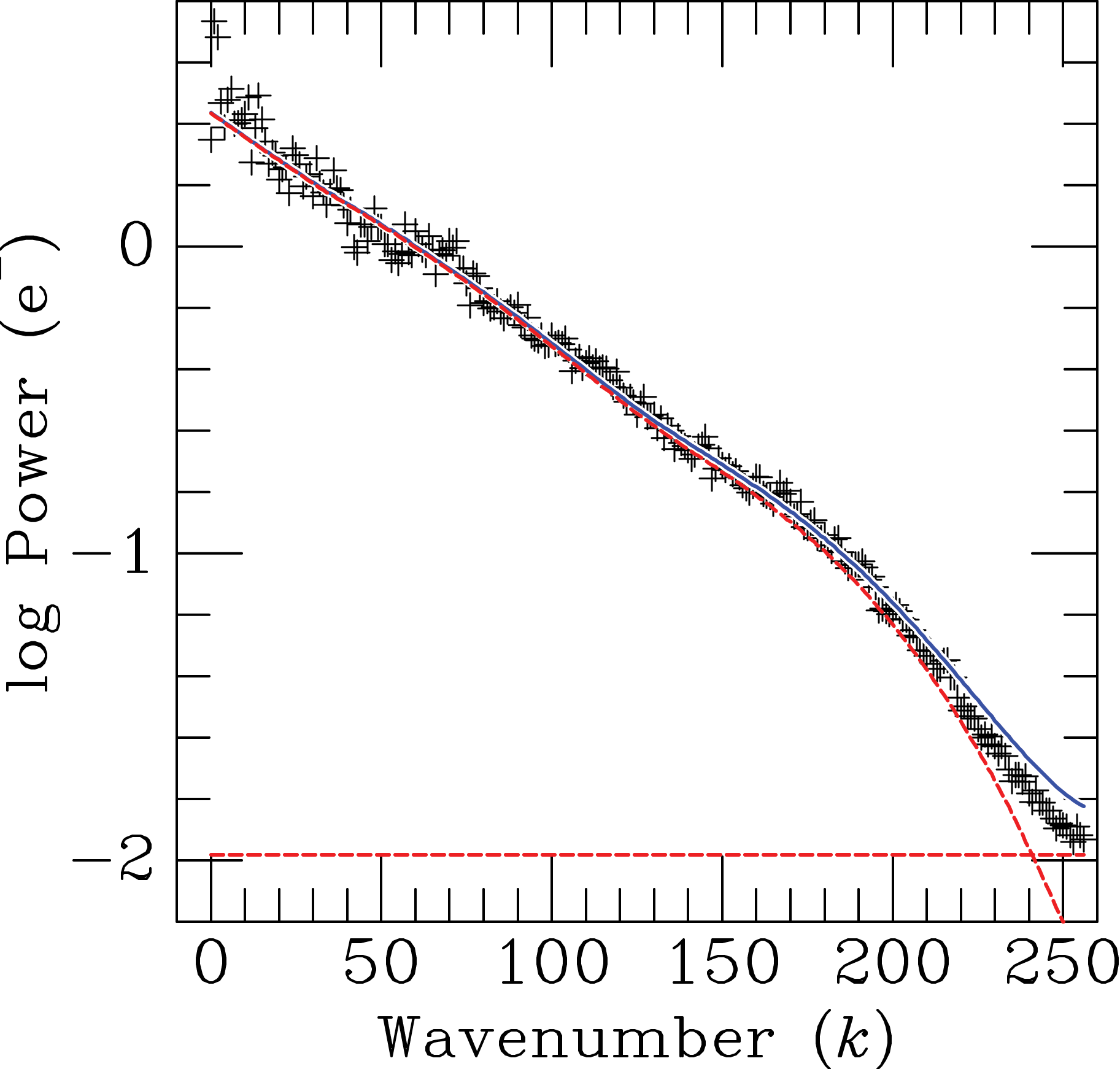}
\caption{Spatial power spectra for NGC\,1399 for the center of the galaxy (B3) in the $F150W$ band and near the peak of the SBF luminosity. The power spectra shows a fit to the PSF power spectrum scaled to match the data and fit in two dimensions, and the constant noise floor (shown with red dotted lines). The blue line shows the excellent fit to the power spectrum over a large range in wavenumber $k$ (x-axis). 
\label{fig:sbfexample}}
\end{figure}

To measure the SBF signal, we first combine the images using integer pixel offsets to eliminate correlations in the noise between pixels. We then determine the background and model the smooth galaxy luminosity profile (Fig.~\ref{fig:sbfexample}). 
The background brightness and galaxy model can be constructed for the different detectors individually and then compared for consistency.  We compute the Fourier spatial power spectrum of the residual image with the galaxy model and background subtracted and extraneous sources like globular cluster masked, and then fit the power spectrum with a scaled power spectrum of the point-spread function. We constructed model PSFs using the WebbPSF software for each NIRCam detector and filter individually, and then convolved the PSFs with Gaussians of various widths in the neighborhood of one pixel for the purpose of ``softening'' the theoretical PSF to match our empirical data, which were combined with integer pixel offsets. An example of the PSF fit to the galaxy power spectrum for NGC~1399 is shown in Fig.~\ref{fig:sbfexample}. The SBF signal-to-noise ratio in this observation is $\sim$250. For these observations at 19~Mpc, the fluctuations are detectable in seconds; detecting the TRGB requires much longer exposures. These SBF measurements are excellent and will provide a high-precision calibration zero point that will be used in the future to measure galaxies to 300~Mpc and beyond with JWST.

\section{Summary and Future Outlook} \label{sec:conc}

In this work, we have presented an analysis of JWST observations of three early-type Fornax Cluster galaxies. With the amazing depth of our data, we are able to perform high-precision measurements of the TRGB in the galaxy halos, allowing us to derive accurate relative distances to these targets, modulo a final setting of the absolute zero-point. Simultaneously, SBF measurements are made in the cores of these galaxies with extremely high precision.

In the near future, we look forward to observations of 11 more early-type galaxy observations from the current program JWST GO--3055, including several massive elliptical galaxies in the Virgo cluster, NGC\,1549 in the Dorado group, and NGC\,3379 in Leo. With a sample of 14 galaxies, we will create a direct linkage between the TRGB and SBF distance scales with better than 2\% accuracy. Previous TRGB measurements did not resolve stars faint enough to clearly detect the TRGB in massive galaxies where SBF can be measured, and the SBF distance scale was calibrated using an indirect linkage to Cepheid variable stars in spiral galaxies. This new TRGB dataset will place future SBF measurements on the secure footing of a geometrical distance zero point with no reliance on different stellar populations or galaxy types. 

Following completion of this project, the next step in building a new, Population II distance ladder will come from SBF observations of a large sample of ${\sim}40$ elliptical galaxies in the Coma Cluster, a JWST program scheduled for Cycle~3~\citep{2024jwst.prop.5989J}.  With a robust calibration of the SBF zero point from TRGB and a broad sampling of the SBF consistency across galaxies of different masses, ages, metallicities, and colors, all at the common distance of the Coma Cluster, SBF will be well calibrated for application at distances of 300~Mpc and beyond with JWST.



\begin{acknowledgments}
This work is based on observations made with the NASA/ESA/CSA JWST. The data were obtained from the Mikulski Archive for Space Telescopes at the Space Telescope Science Institute, which is operated by the Association of Universities for Research in Astronomy, Inc., under NASA contract NAS 5-03127 for JWST. These observations are associated with program \#3055. The specific observations analyzed can be accessed via \dataset[10.17909/c25f-g512]{https://archive.stsci.edu/doi/resolve/resolve.html?doi=10.17909/c25f-g512}. 

G.S.A.{} thanks Rachael Beaton and Adam Riess for useful discussions. G.S.A.{}, Y.C.{}, and J.B.J.\ acknowledge financial support from JWST GO–3055.
D.I.M.{} and L.N.M.{} are supported by the grant \textnumero~075--15--2022--262 (13.MNPMU.21.0003) of the Ministry of Science and Higher Education of the Russian Federation.

This research made use of the NASA Astrophysics Data System. 
This research also used the NASA/IPAC Extragalactic Database (NED), which is funded by the National Aeronautics and Space Administration and operated by the California Institute of Technology.

The Digitized Sky Surveys were produced at the Space Telescope Science Institute under U.S. Government grant NAG W-2166. The images of these surveys are based on photographic data obtained using the Oschin Schmidt Telescope on Palomar Mountain and the UK Schmidt Telescope. The plates were processed into the present compressed digital form with the permission of these institutions.

\end{acknowledgments}


\bibliography{paper}{}
\bibliographystyle{aasjournal}



\end{document}